\documentclass{article}

\usepackage{microtype}
\usepackage{graphicx}
\usepackage{subfig}
\usepackage{booktabs} 
\usepackage{interval}

\usepackage{hyperref}

\usepackage[accepted]{icmlfake}
\usepackage{amsmath,amssymb,amsfonts}
\usepackage[font=small,labelfont=bf]{caption}

\icmltitlerunning{Generative adversarial networks for brain signals}

\begin{document}

\twocolumn[
\icmltitle{EEG-GAN: Generative adversarial networks for electroencephalograhic (EEG) brain signals}



\icmlsetsymbol{equal}{*}

\begin{icmlauthorlist}
\icmlauthor{Kay Gregor Hartmann}{tnt}
\icmlauthor{Robin Tibor Schirrmeister}{tnt}
\icmlauthor{Tonio Ball}{tnt}
\end{icmlauthorlist}

\icmlaffiliation{tnt}{Translational Neurotechnology Lab, Medical Center, University of Freiburg, Freiburg i.Br., Germany}

\icmlcorrespondingauthor{Kay Hartmann}{kg.hartma@gmail.com}

\icmlkeywords{Machine Learning, GAN, EEG, WGAN, Inception score, Frechet Inception Distance}

\vskip 0.3in]



\printAffiliationsAndNotice{}  

\begin{abstract}
Generative adversarial networks (GANs) are recently highly successful in generative applications involving images and start being applied to time series data. Here we describe EEG-GAN as a framework to generate electroencephalographic (EEG) brain signals. We introduce a modification to the improved training of Wasserstein GANs to stabilize training and investigate a range of architectural choices critical for time series generation (most notably up- and down-sampling). For evaluation we consider and compare different metrics such as Inception score, Frechet inception distance and sliced Wasserstein distance, together showing that our EEG-GAN framework generated naturalistic EEG examples. It thus opens up a range of new generative application scenarios in the neuroscientific and neurological context, such as data augmentation in brain-computer interfacing tasks, EEG super-sampling, or restoration of corrupted data segments. The possibility to generate signals of a certain class and/or with specific properties may also open a new avenue for research into the underlying structure of brain signals.
\end{abstract}

\section{Introduction}
While large parts of machine learning deal with the decoding of information from real-world data such as in classification tasks, there is also the recently very active field of how to  generate such real-world data through implicit generative models. Generating artificial data could for example be used for data augmentation by producing natural looking samples that are not included in the original data set and thereby artificially increase training data with unseen samples. Additionally, the possibility to produce natural looking samples with certain properties, and the investigation of the models creating them, can be a useful tool for understanding the original data distribution used for training the GAN

One recently proposed framework for the generation of artificial data are generative adversarial networks \cite{Goodfellow2014} which showed groundbreaking results for the generation of artificial images. Originally, vanilla GANs suffered heavily from training instability and were restricted to low resolution images. A lot of advancement in regard to stability and the quality of the generated images has been made with different regularization methods \cite{Mao2016,Arjovsky2017,Gulrajani2017,Kodali2017} and by progressively increasing the image resolution during training \cite{Karras2017}. GANs also allow the intentional manipulation of specific properties in generated samples \cite{Radford2015} and therefore could prove to be a useful tool in understanding the original data distribution used for training the GAN.

GANs have mainly been developed and applied to the generation of images and only a few studies investigating time series were conducted; recently they showed promising results for the generation of artificial audio \cite{Donahue2018}. The generation of artificial EEG signals would have applications in many different areas dealing with decoding and understanding brain signals, but to our best knowledge no research regarding the generation of raw EEG signals with GANs has been published at this time.

In this work, we apply the GAN framework to the generation of artificial EEG signals. Though the generation of time-series data is often approached with autoregressive models (e.g. WaveGAN by \citet{Oord2016}), we deliberately chose regular convolutional neural networks - on the one hand because most of GAN studies use the DCGAN \cite{Radford2015} architecture which is based on CNNs, on the other hand because the local and hierarchical structure of CNNs may allow for better interpretability \cite{Sturm2016,Kindermans2017,Schirrmeister2017,Hartmann2018} that is particularly important for brain singals in a neuroscientific or clinical context.
To generate naturalistic samples of EEG data, we propose an improvement to the Wasserstein GAN training showing increased training stability. Furthermore, we compare different evaluation metrics and discuss methodological and architectural choices of the network that delivered the best results in this study.

\section{Methods}
\subsection{GAN background and improvement}
The GAN framework consists of two opposing networks trying to outplay each other \cite{Goodfellow2014}. The first network, the discriminator, is trained to distinguish between real and fake input data. The second network, the generator, takes a latent noise variable z as input and tries to generate fake samples that are not recognized as fake by the discriminator. This results in a minimax game in which the generator is forced by the discriminator to produce ever better samples.

One big drawback of GANs is the notorious instability of the discriminator during training. The discriminator might collapse into only recognizing few and narrow modes of the input distribution as real, which drives the generator to produce only a limited amount of different outputs. Mode collapse is very problematic for training GANs and is the subject of various works \cite{Mao2016,Arjovsky2017,Gulrajani2017,Kodali2017}.

Wasserstein GANs and their improved version proposed by \citet{Arjovsky2017} show promising advances for training stability. The original GAN framework tries to minimize the Jensen-Shannon (JS) divergence between the real data distribution $\mathbb{P}_r$ and fake data distribution $\mathbb{P}_{\theta}$ \cite{Goodfellow2014}. If the discriminator is trained to optimality this may lead to the problem of vanishing gradients for the generator \cite{Arjovsky2017}. \citet{Arjovsky2017} proposed to  to minimize the Wasserstein distance between the distributions instead of the JS-divergence . This leads the discriminator (now called critic) to maximize the difference
\begin{equation}
\begin{split}
\begin{aligned}
    \tilde{W}(\mathbb{P}_r,\mathbb{P}_{\theta}) = E_{x_r\sim \mathbb{P}_r}[D(x_r)]-E_{x_f\sim \mathbb{P}_{\theta}}[D(x_f)]
\end{aligned}
\end{split}
\end{equation}
and the generator to maximize $E_{x_f\sim \mathbb{P}_{\theta}}[D(x_f)]$. They showed that the critic provides a useful gradient for the generator everywhere if $D(x)$ is K-Lipschitz. In their original paper, they enforced Lipschitz continuity by clipping the weights of the discriminator to an interval $\interval{-c}{c}$ (WGAN-clip), but later derived a more elegant solution by adding a gradient penalty term
\begin{equation}
\begin{split}
\begin{aligned}
    P_{2}(\mathbb{P}_{\hat{x}}) = \lambda\cdot E_{\hat{x} \sim \mathbb{P}_{\hat{x}}}[(||\nabla_{\hat{x}} D(\hat{x})||_2-1)^2]
\end{aligned}
\end{split}
\end{equation}
with $\mathbb{P}_{\hat{x}}$ containing the points lying on a straight line between real and generated samples, to the critic loss \cite{Gulrajani2017}.

The choice of $\lambda$ is crucial when training WGANs with gradient penalty (WGAN-GP). If $\lambda$ is chosen too high, the penalty term can easily dominate the distance term. The other way around, if $\lambda$ is chosen too small, the lipschitz continuity is not sufficiently enforced. We noticed that a good choice for $\lambda$ heavily depends on the distance between $\mathbb{P}_r$ and $\mathbb{P}_{\theta}$. If the distance is high, $\lambda$ has to be chosen accordingly high. If they are close, $\lambda$ has to be chosen accordingly low. However, during training the generator learns to approximate $\mathbb{P}_r$. This leads to a decrease of the distance between $\mathbb{P}_r$ and $\mathbb{P}_{\theta}$, whereas $\lambda$ stays constant.

\begin{figure}[ht]
\centering
    \subfloat[]{%
       \includegraphics[width=0.48\columnwidth]{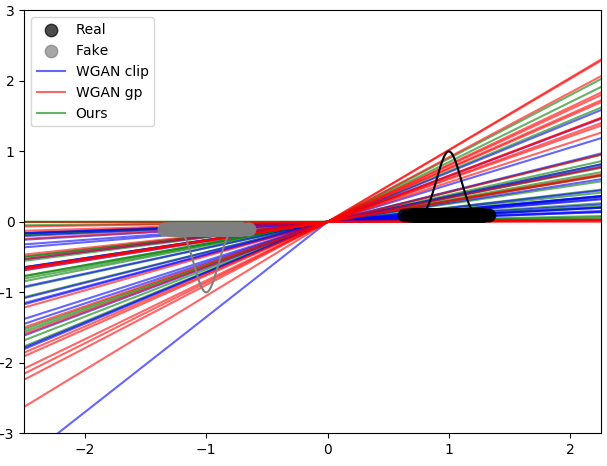}
       \label{fig:wgan1}
     }
     \hfill
     \subfloat[]{%
       \includegraphics[width=0.48\columnwidth]{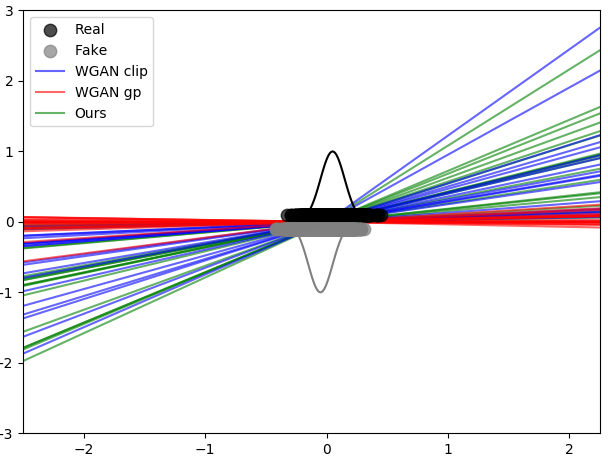}
       \label{fig:wgan2}
     }
\caption{WGAN critics trained to optimality for two distributions being \textbf{(a)} distant and  \textbf{(b)} near.}
\label{fig:wgan}
\vskip -0.1in
\end{figure}

Figure \ref{fig:wgan1} shows WGAN-clip and WGAN-GP critics trained to distinguish between two normal distributions. Parameters $c$ and $\lambda$ were chosen such that the critics provide a useful gradient and $||\nabla_{\hat{x}} D(\hat{x})||_2<=1$ for most critics . Figure \ref{fig:wgan2} again shows critics with the same parameter setting trained to distinguish between two distributions, but now with decreased distance between them. This simulates the approximation of $\mathbb{P}_{\theta}$ to $\mathbb{P}_r$ by the generator. WGAN-GP critics noticeably collapse into vanishing gradients and in several cases displaying $E_{x_f\sim \mathbb{P}_{\theta}}[D(x_f)]>E_{x_r\sim \mathbb{P}_r}[D(x_r)]$. WGAN with weight clipping remains stable, as its only objective is to maximize the critic difference regularized only by limiting network weights. Limiting weights, however, leads to an undesired convergence of network parameters to those limits \cite{Gulrajani2017}.

Therefore, we propose an improvement to WGAN-GP by gradually relaxing the gradient constraint. Instead of only weighting the penalty term with $\lambda$, we also scale it by the current critic difference $\tilde{W}(\mathbb{P}_r,\mathbb{P}_{\theta})$. Thereby, the penalty term is only heavily enforced if the first objective of the critic to distinguish between $\mathbb{P}_r$ and $\mathbb{P}_{\theta}$ is met and additionally $\lambda$ is scaled down for decreasing distribution distances. Additionally, we will not use the two-sided penalty $P_{2}(\mathbb{P}_{\hat{x}})$ recommended by \citet{Gulrajani2017}, but the one-sided penalty
\begin{equation}
\begin{split}
\begin{aligned}
    P_{1}(\mathbb{P}_{\hat{x}}) = \lambda\cdot E_{\hat{x} \sim \mathbb{P}_{\hat{x}}}[\max(0,||\nabla_{\hat{x}} D(\hat{x})||_2-1)^2]).
\end{aligned}
\end{split}
\end{equation}
They did not state a specific reason to choose the two-sided penalty over the one-sided penalty, but preferred it from empirical results. The resulting loss function for the critic then becomes:
\begin{equation}
\begin{split}
\begin{aligned}
    L_c = -\tilde{W}(\mathbb{P}_r,\mathbb{P}_{\theta})+\max(0,\tilde{W}(\mathbb{P}_r,\mathbb{P}_{\theta}))\cdot P_{1}(\mathbb{P}_{\hat{x}}).
\end{aligned}
\end{split}
\end{equation}
Critics trained with this loss exhibit stable gradients for distributions with decreasing distance (Figure \ref{fig:wgan}).

\subsection{Training and architecture choices}
We trained our networks according to the setup described in \citet{Karras2017} They showed that the generated image quality can be enhanced by training the network progressively with increasing resolution. Accordingly, we start at a resolution of 24 time samples and increase the resolution by factor 2 over 6 steps to arrive at 768 samples. Factor 2 introduced the least frequency artifacts and led to the best results. We included additional techniques from \citet{Karras2017} such as minibatch standard deviation, equalized learning rate and pixel normalization. Instead of their proposed additional penalty term $\epsilon\cdot E_{x_r\sim \mathbb{P}_r}[D(x_r)^2]$ to keep the critic from drifting too far away from 0, we instead use $\epsilon\cdot(E_{x_r\sim \mathbb{P}_r}[D(x_r)]+E_{x_f\sim \mathbb{P}_{\theta}}[D(x_f)])^2$  with $\epsilon=0.001$ to keep the critic centered at 0.

In opposition to \citet{Karras2017}, we do not train the critic and generator equally, but train the critic until optimality first (by 5 critic iterations, as originally proposed by \citet{Arjovsky2017}). We set $\lambda=10$, as originally recommended by \citet{Gulrajani2017} Each resolution stage is trained for 2000 epochs (which equals to 876.000 signal showings), with an additional 2000 epochs for fading in each stage. The networks are trained using the ADAM optimizer \cite{Kingma2014} with $lr=0.001$, $\beta_1=0$ and $\beta_2=0.99$. Latent variables $z$ for the generator are sampled from $\mathcal{N}(0,1)$.

Our network architecture can be seen in Table \ref{tabl:architecture}. Each upsampling block in the generator consists of an upsampling layer followed by 2 convolution layers of size 9. Similarly, each critic block consists of 2 convolution layers followed by 1 downsampling layer. For downsampling we used average pooling and strided convolutions with a size and stride of 2. We use leaky ReLUs in the critic and generator to avoid sparse gradients.
\begin{table}[t!]
\caption{Network architecture}
\resizebox{\columnwidth}{!}{%
\subfloat[Generator]{
\begin{tabular}[b]{|lcc|}
\hline
\textbf{Layer} & \textbf{Act./Norm.} & \textbf{Output shape} \\ \hline
Latent vector                        & -                   & 200 x 1               \\
Linear                               & LReLU               & 50 x 12               \\ \hline
Upsample                             & -                   & 50 x 24               \\
Conv 9                               & LReLU/PN            & 50 x 24               \\
Conv 9                               & LReLU/PN            & 50 x 24               \\ \hline
Upsample                             & -                   & 50 x 48               \\
Conv 9                               & LReLU/PN            & 50 x 48               \\
Conv 9                               & LReLU/PN            & 50 x 48               \\ \hline
Upsample                             & -                   & 50 x 96               \\
Conv 9                               & LReLU/PN            & 50 x 96               \\
Conv 9                               & LReLU/PN            & 50 x 96               \\ \hline
Upsample                             & -                   & 50 x 192              \\
Conv 9                               & LReLU/PN            & 50 x 192              \\
Conv 9                               & LReLU/PN            & 50 x 192              \\ \hline
Upsample                             & -                   & 50 x 384              \\
Conv 9                               & LReLU/PN            & 50 x 384              \\
Conv 9                               & LReLU/PN            & 50 x 384              \\ \hline
Upsample                             & -                   & 50 x 768              \\
Conv 9                               & LReLU/PN            & 50 x 768              \\
Conv 9                               & LReLU/PN            & 50 x 768              \\ \hline
Conv 1                               & -                   & 1 x 768               \\ \hline
\end{tabular}}
\subfloat[Critic]{
\begin{tabular}[b]{|lcc|}
\hline
\tiny
\textbf{Layer} & \textbf{Act.} & \textbf{Output shape} \\ \hline
Input signal   & -             & 1 x 768               \\
Conv 1         & LReLU         & 50 x 768              \\ \hline
Conv 9         & LReLU         & 50 x 768              \\
Conv 9         & LReLU         & 50 x 768              \\
Downsample     & -             & 50 x 384              \\ \hline
Conv 9         & LReLU         & 50 x 384              \\
Conv 9         & LReLU         & 50 x 384              \\
Downsample     & -             & 50 x 192              \\ \hline
Conv 9         & LReLU         & 50 x 192              \\
Conv 9         & LReLU         & 50 x 192              \\
Downsample     & -             & 50 x 96               \\ \hline
Conv 9         & LReLU         & 50 x 96               \\
Conv 9         & LReLU         & 50 x 96               \\
Downsample     & -             & 50 x 48               \\ \hline
Conv 9         & LReLU         & 50 x 48               \\
Conv 9         & LReLU         & 50 x 48               \\
Downsample     & -             & 50 x 24               \\ \hline
Conv 9         & LReLU         & 50 x 24               \\
Conv 9         & LReLU         & 50 x 24               \\
Downsample     & -             & 50 x 12               \\ \hline
Linear         & -             & 1 x 1                 \\ \hline
\end{tabular}}}
\label{tabl:architecture}
\vskip -0.2in
\end{table}

\begin{figure}[t]
\begin{center}
\includegraphics[width=\columnwidth]{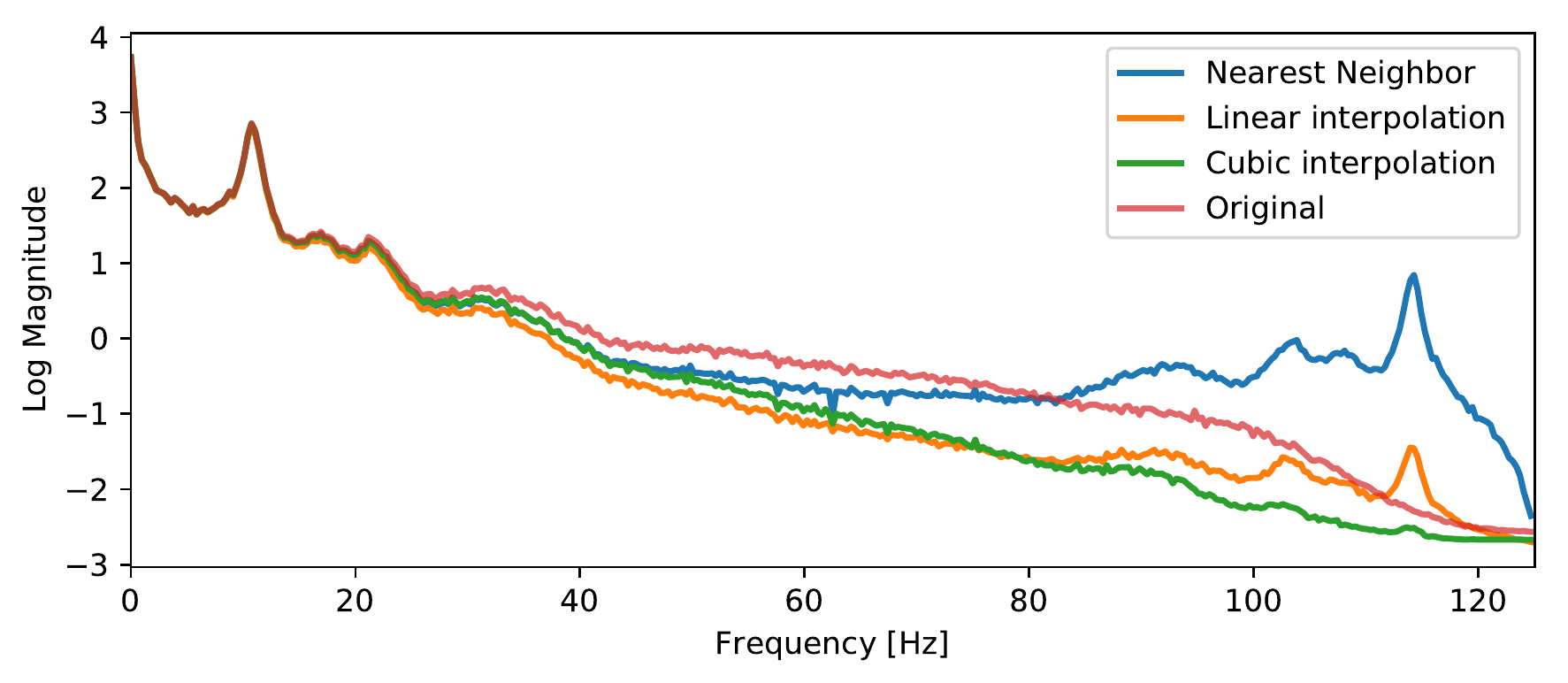}
\vskip -0.1in
\caption{Mean frequency spectra of EEG signals after upsampling visualizing aliasing artifacts in high frequencies. The original signals were first downsampled by average pooling and then upsampled by different upsampling methods.}
\label{fig:upsample}
\end{center}
\vskip -0.2in
\end{figure}

For upsampling we compared nearest-neighbor upsampling and linear and cubic interpolation. As stated by \citet{Donahue2018}, upsampling always introduces aliasing frequency artifacts. Whereas they argue that those artifacts may be necessary to produce fine-grained details, we believe that it unnecessarily complicates training for the generator, at least in the case of EEG signals. Nearest-neighbor upsampling introduces strong high-frequency artifacts, while linear or cubic interpolation lead to much weaker artifacts (Figure \ref{fig:upsample}). We argue this is favorable, because we do not want the generator to filter out artifacts after upsampling, but expect the generation of additional high-frequency features.

\subsection{Evaluation metrics}
\subsubsection{Inception score}
Evaluation of the distribution of the samples generated by of the generator is an ongoing challenge. Visual inspection of generated samples can be helpful to identify obvious failures and mode collapse, but fail to give any quantitative information about the variance of generated samples and how similar they are to the training data. A common approach to give information about the quality of the trained generator is to use the inception score (IS) \cite{Salimans2016a}. To calculate the inception score, a classifier is trained on the training data and used to determine the entropy of the conditional label distribution of generated samples, which should be low, and the entropy of their marginals, which should be high. Though the inception score was shown to be well-correlated with author annotations, it fails to give useful information about the quality of the generator. The inception score is highly sensitive to noise and is not able to detect mode collapse, as it solely relies on the final probabilities of the classifier.  We used a pretrained Deep4 model \cite{Schirrmeister2017} as a replacement for the inception model for EEG data.

\subsubsection{Frechet inception distance}
The Frechet inception distance (FID) \cite{Heusel2017} aims to give a better evaluation of the quality of generated samples and is a proper distance. Similarly to the inception score, the FID also uses a trained classifier. But instead of simply evaluating the distribution of class probabilities of generated samples, the FID compares the values in the embedding layer (i.e., the layer before  the final classification layer) for real and generated samples. The Frechet distance is used to calculate the Wasserstein2 distance between the distributions of values in the embedding layer for real and fake samples, under the assumption that they follow a multivariate Gaussian distribution. The FID has been shown by Heusel et al. to be consistent with human judgment, and, in contrast to the inception score, more robust to noise, giving information about the quality of the generated samples and to be sensitive to mode collapse. However, it is also not able to detect overfitting of the generator to training samples.

\subsubsection{Euclidean distance}
The euclidean distance can be used to evaluate how similar generated samples are to the training data. By comparing the distances between generated and real samples, we can investigate of the generator simply replicates samples from the training set or produces something unseen, a property often not evaluated for generative models, but especially important for us, since we have a lower amount of training samples compared to very large image datasets. Optimally, the distribution of minimal distances between real and fake samples (ED\textsubscript{min}) should be equivalent to the distribution of minimal distances between only real samples with others than themselves.

\subsubsection{Sliced Wasserstein distance}
The Wasserstein distance describes the cost of transforming one distribution to another using under a given cost function (see \cite{Peyre2018} for a more detailed explanation and an overview). The sliced Wasserstein distance (SWD) is an approximation to the Wasserstein distance using 1d projections. It approximates the Wasserstein distance by computing Wasserstein distances between all 1d-projections (slices) of the two distributions. This has the advantage of a closed-form solution and corresponding fast computation for the 1d-case. In practice, the sliced Wasserstein distance is itself  approximated by using a finite set of random 1d-projections \cite{Rabin2012}. A low sliced Wasserstein distance indicates that the two distributions are similar in their appearance and variation of samples.

\section{Data}
The EEG signals we will use for training stem from a simple motor task in which the subjects were instructed to either rest or move the left hand. The signals were recorded with a 128-electrode EEG system and downsampled to 250 Hz. The subject showed characteristic spectral information for left hand movement at electrode channel FCC4h in the alpha, beta and high gamma range. Here, we will only use channel FCC4h for training the GAN. The dataset was scaled to $\interval{-1}{1}$ by subtracting the mean and then dividing by the maximum absolute value. Overall the dataset contains 438 signals, from which 286 will be used as training data for the inception classifier, 72 as validation and 80 as test set. All 438 signals will be used for training the GAN.

\begin{table}[ht]
\centering
\begin{tabular}{|llrrrr|}
\hline
\textbf{\#} & \textbf{Model}                                             & \multicolumn{1}{l}{\textbf{IS}} & \multicolumn{1}{l}{\textbf{FID}} & \multicolumn{1}{l}{\textbf{ED\textsubscript{min}}} & \multicolumn{1}{l|}{\textbf{SWD}} \\ \hline
1           & AVG-NN                                                     & 1.361                           & \textbf{9.523}                   & \textbf{-0.056}                    & \textit{0.102}                             \\
2           & CONV-NN                                                    & 1.297                           & 16.755                           & -0.121                             & 0.084                             \\
3           & CONV-LIN                                                   & \textbf{1.363}                  & 11.854                           & -0.252                             & 0.086                             \\
4           & CONV-CUB                                                   & \textit{1.292}                  & \textit{33.765}                  & \textit{-0.375}                    & \textbf{0.078}                    \\
\hline
5           & \begin{tabular}[c]{@{}l@{}}WGAN-GP\\ CONV-CUB\end{tabular} & 1.281                           & 120.854                          & +0.034                             & 0.309                             \\
\hline
            & Real                                                       & 1.555                           & 0.                               & 4.653                              & 0.                                \\
            & Noise                                                      & 1.049                           & 614.782                          & +1.061                             & 0.155                             \\ \hline
\end{tabular}
\caption{Results for GANs with different architectures. AVG denotes average pooling, CONV strided convolution as downsampling. NN denotes nearest-neighbor upsampling, LIN linear and CUB cubic interpolation. All models except WGAN-GP were trained with our method. WGAN-GP collapsed during training. Best scores are marked by bold, worst scores by italic fonts.}
\label{tabl:distances}
\vskip -0.1in
\end{table}
\begin{figure*}[ht]
\centering
\includegraphics[width=\textwidth]{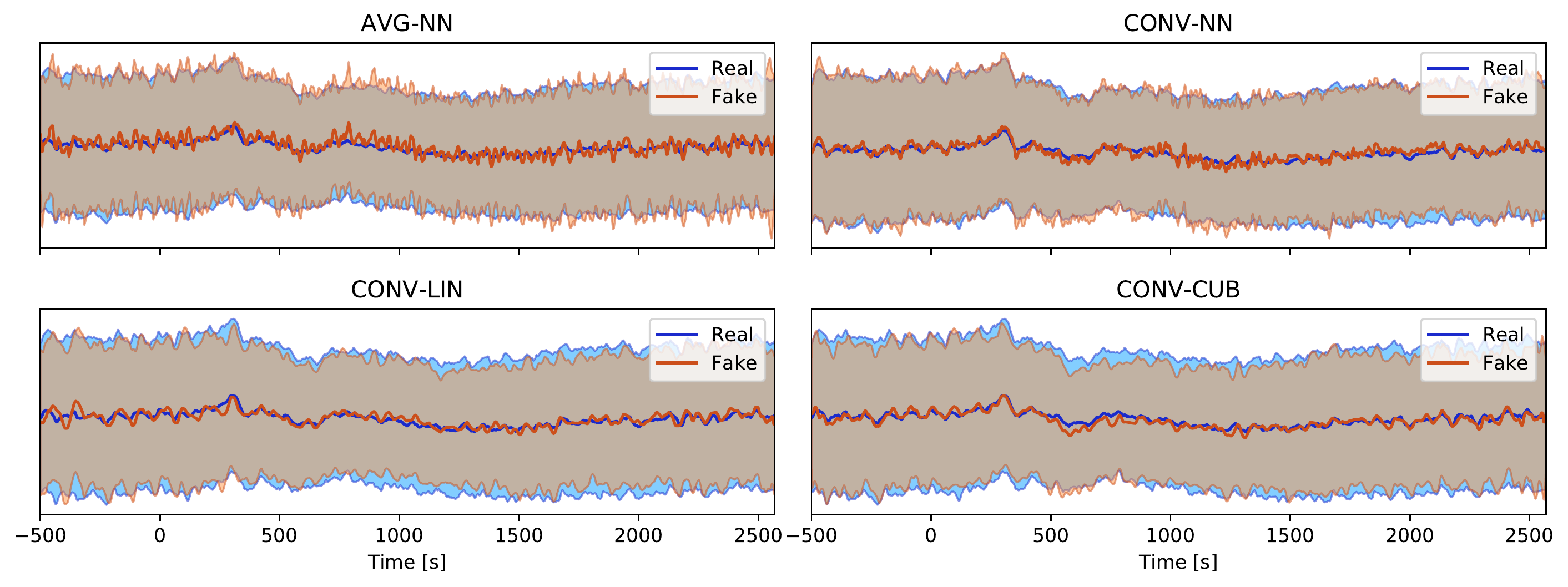}
\caption{Comparison of the distribution of values at each time point between real signals and signals created by different architectures.}
\label{fig:time}
\end{figure*}
\section{Results}
\subsection{Distance results}
Table \ref{tabl:distances} shows metric results for different architectures trained with our method (1-4) and the original WGAN-GP (5). AVG denotes downsampling by average pooling, CONV by strided convolution. For upsampling, NN denotes nearest neighbor, LIN linear interpolation and CUB cubic interpolation. The test accuracy of the inception classifier used for calculating the inception score and Frechet inception distance was 91.25\%. Scores for real data and noise (sampled from a normal distribution with mean and variance of the real data) are listed for comparison.

From visual inspection and the FID and sliced Wasserstein distance, it was clear that the WGAN-GP model collapsed, though we have to note that we neither performed any hyperparameter search nor conducted multiple runs to find a working model. However, we neither did this for the models trained with our method and, though they varied in performance, none of them collapsed.The IS gave no strong evidence for the collapse of the model.

For the models trained with our method, different architectures performed best for different metrics. CONV-LIN performed best for IS with AVG-NN being a close second. AVG-NN did perform best and CONV-CUB clearly worst for FID. EDmin was closest to the real EDmin for AVG-NN, again with CONV-CUB being the worst. For SWD however, CONV-CUB was clearly the best and AVG-NN clearly the worst performing model. Overall, CONV-CUB was the worst performing model for all metrics except SWD.

\subsection{Visual inspection}
\subsubsection{Time samples}
Figure \ref{fig:time} shows the mean and standard deviation from signals created by the 4 architectures trained by our model compared to the real signals. The similarity of samples at each time point increases from architecture 1 to 4.. Whereas AVG-NN shows a clear deviation of the generated sample distributions from real data, CONV-CUB shows a very good fit.

\subsubsection{Frequency spectra}
\begin{figure}[ht]
\begin{center}
\includegraphics[width=\columnwidth]{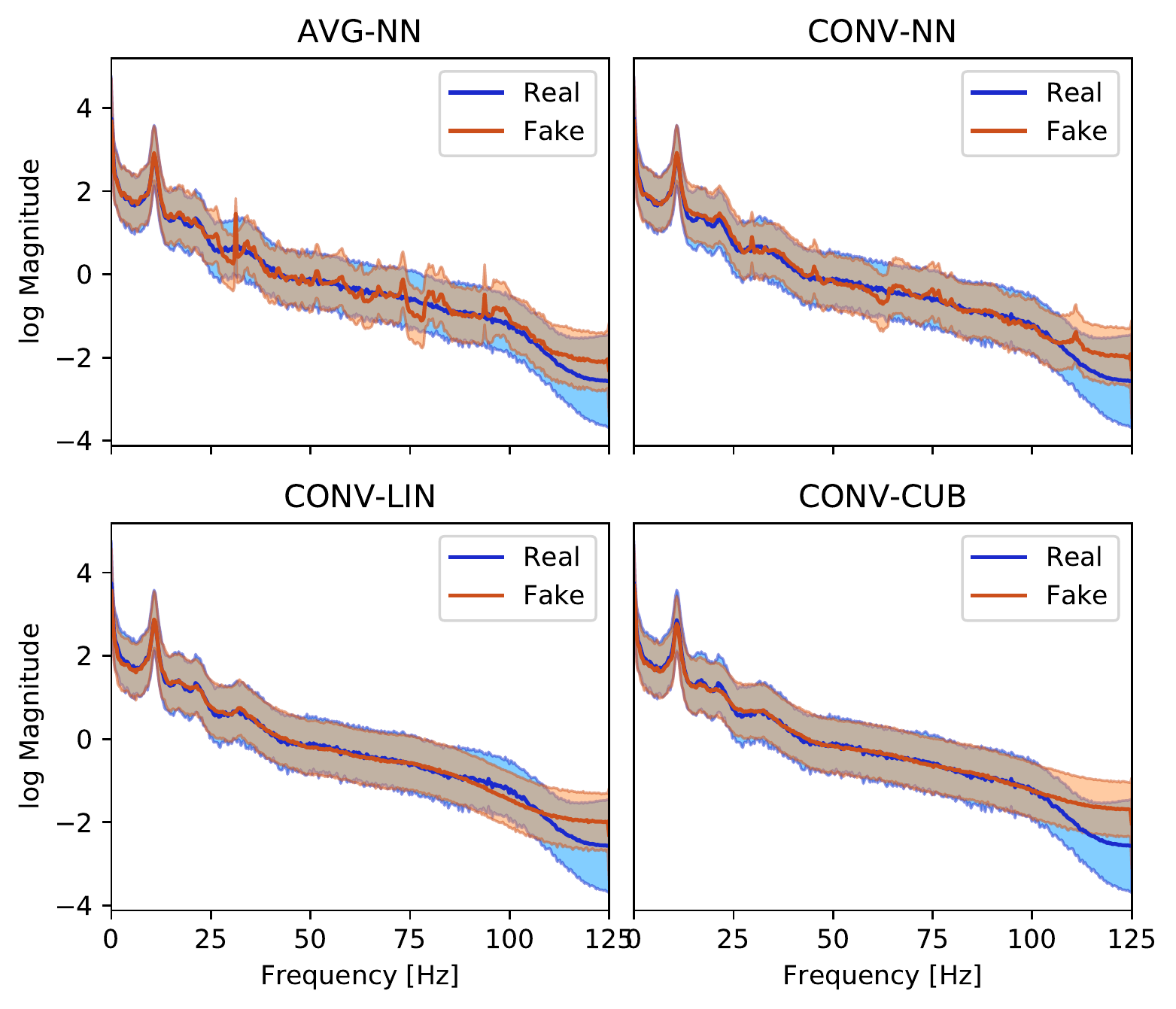}
\vskip -0.1in
\caption{Comparison of the distribution of frequency spectra real signals and signals created by different architectures.}
\label{fig:freqs}
\end{center}
\vskip -0.2in
\end{figure}

Similarly, Figure \ref{fig:freqs} shows the comparison of frequency-resolved spectral power distributions. AVG-NN and CONV-NN show deviations from the real spectrum even in low frequencies, whereas CONV-LIN and CONV-CUB again show a good fit. It can be argued that CONV-LIN better fits low frequencies, whereas CONV-LIN better fits high frequencies. No model managed to properly fit frequencies higher than 100 Hz (which however have very low power).

\subsubsection{Generated samples}
\begin{figure}[t]
\centering
    \subfloat[\label{fig:sig1}]{%
       \includegraphics[width=0.48\columnwidth]{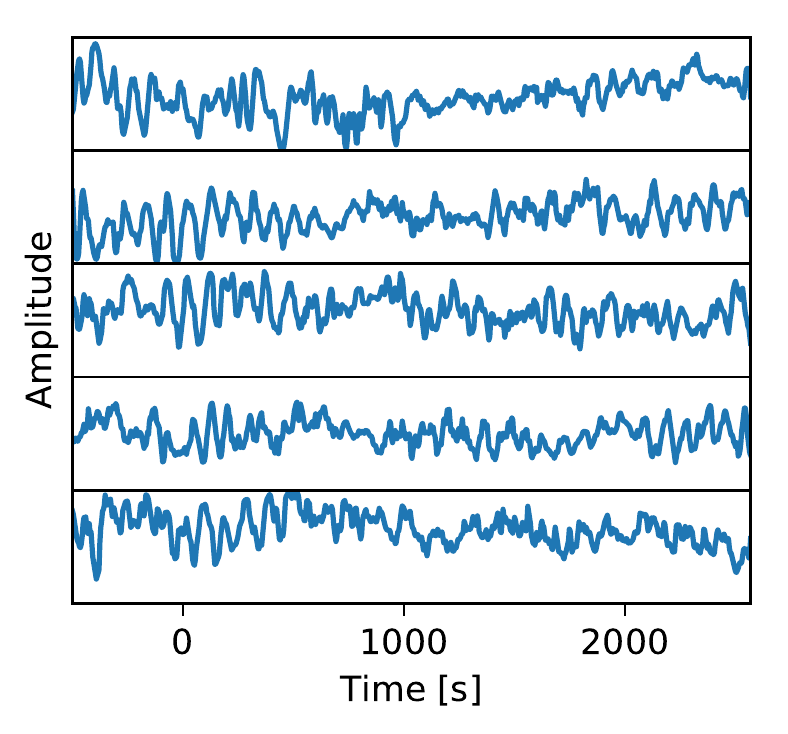}
     }
     \hfill
     \subfloat[\label{fig:sig2}]{%
       \includegraphics[width=0.48\columnwidth]{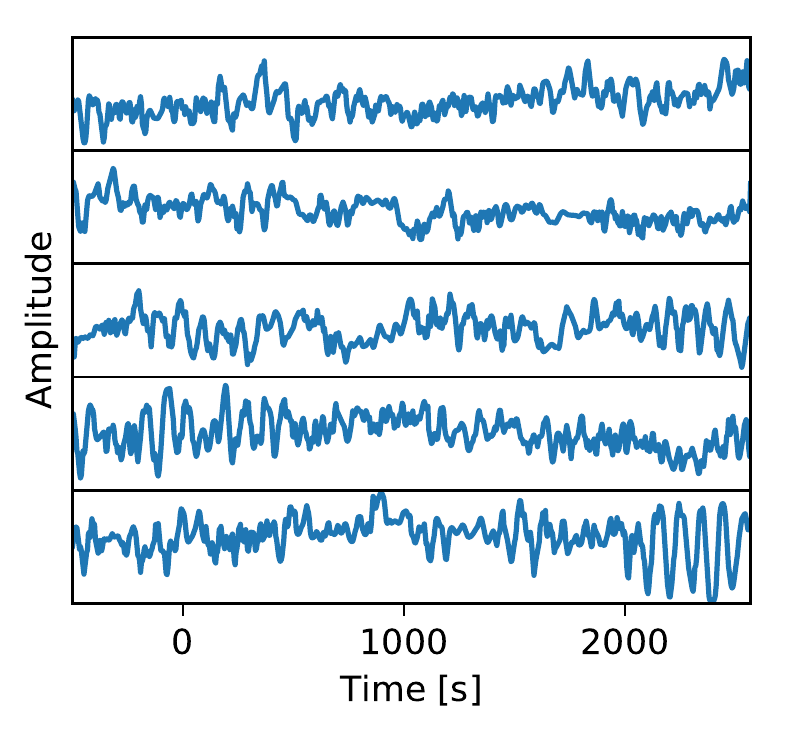}
     }
\vskip -0.1in
\caption{Signals created by the \textbf{(a)} AVG-NN and \textbf{(b)} CONV-CUB models.}
\label{fig:sig}
\vskip -0.1in
\end{figure}

Figure \ref{fig:sig} shows random samples generated by AVG-NN and CONV-CUB. Both models appear to generate visually sound signals. A notable difference between the signals is that CONV-CUB decently often creates signals containing only weakly oscillating sequences. Those sequences were highly indicative for fake signals when surveyed by visual comparison to real signals.

\subsection{Class-specific properties}
\begin{figure}[t]
\begin{center}
\includegraphics[width=\columnwidth]{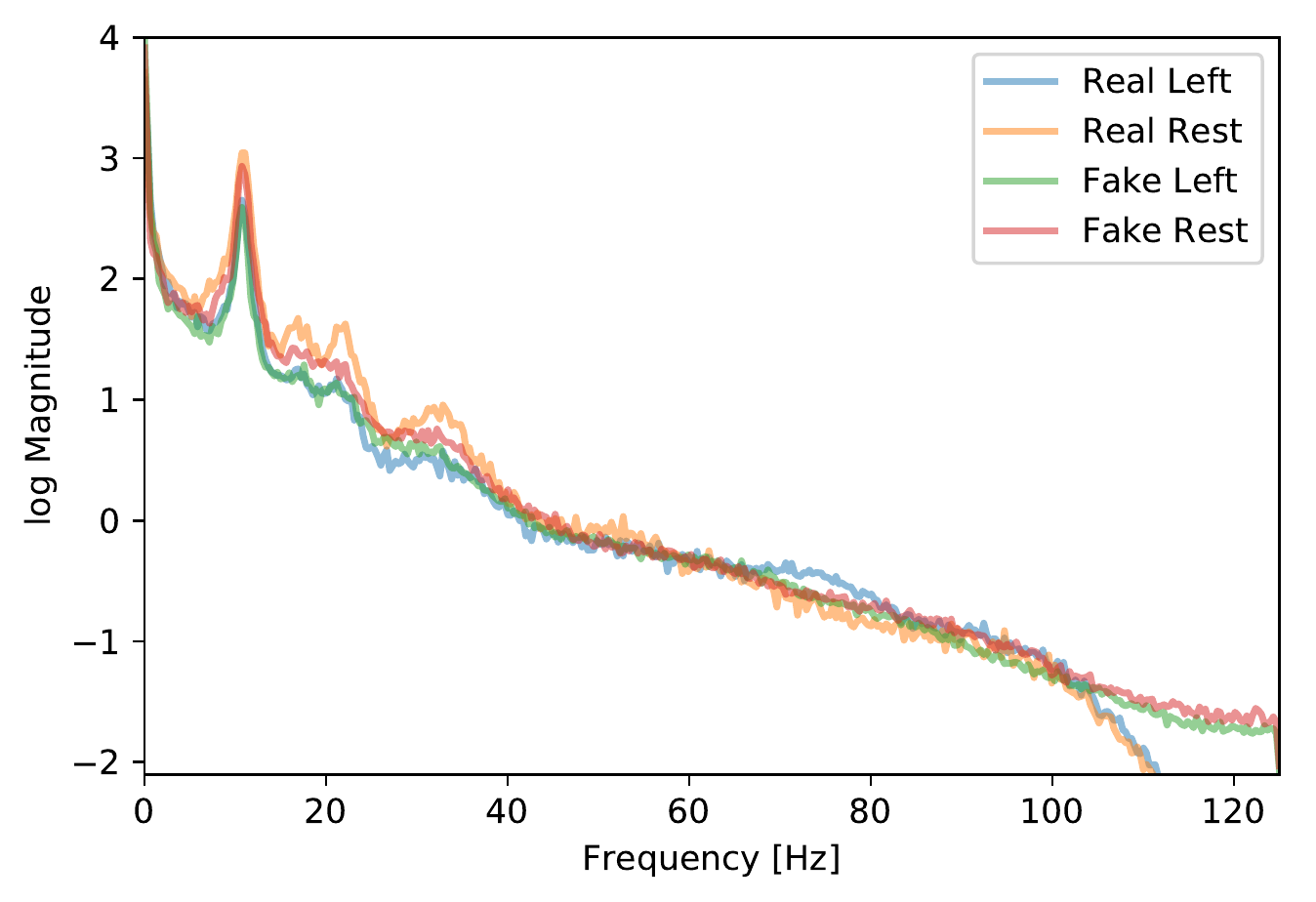}
\vskip -0.1in
\caption{Comparison of mean frequency spectra of real and generated signals that were either classified as either resting activity or left hand movement.}
\label{fig:classprops}
\end{center}
\vskip -0.2in
\end{figure}

To investigate class specific properties of signals generated by CONV-CUB, we used the inception classifier to determine signals that were classified to belong to either class with \textgreater90\% accuracy. Similarly, we used the same approach to determine real signals that exhibited a probability of \textgreater90\% for either class. A comparison of the respective frequency spectra is shown in Figure \ref{fig:classprops}. Generated signals classified as left hand movement match the decrease of alpha and beta activity in the real signals up to around 40 Hz. Generated left hand signals though did not express the increase of high gamma activity present in real left hand signals.

\section{Discussion \& Conclusion}
With this work, we showed that it is possible to generate artificial EEG signals with a generative adversarial network. With a modification of the improved WGAN training \cite{Gulrajani2017}, we were able to progressively train a GAN to produce artificial signals in a stable fashion that strongly resemble single-channel real EEG signals in the time and frequency domain. 

We evaluated the up- and down-sides of several choices for up- and downsampling by comparing them with various metrics. In our case, the inception score (IS) \cite{Salimans2016a} did not give meaningful information about the quality of signals generated by a model. Additionally we observed that models with the lowest Frechet inception distances (FID) \cite{Heusel2017} did not necessarily produce signals with spatial and spectral properties similar to the real input samples. The model that produced the most naturally distributed signals according to spatial and spectral properties was assigned the worst FID. Therefore optimization of GANs used for EEG towards good IS or FID could lead to the production of signal distributions wrongly believed to be similar to real data. The model expressing the most natural looking spatial and spectral distributions had the best sliced Wasserstein distance (SWD).  A low Euclidean distance suggests a preference of the generator towards specific training samples, though in our case it was not as low that it indicates the simple reproduction of training samples. Overall, no single metric gave sufficient information about the quality of a model, but the combination of FID, SWD and ED gave a good idea about its possible overall properties. Therefore we do not recommend any single metric, but encourage the use of several metrics with different advantages and disadvantages.
encourage the use of several metrics with different advantages and disadvantages.

\section{Outlook}
With the first step for the generation of artificial EEG signals done, there are now many open possibilities for further investigations. Of course, the next step would be to not only generate single channel signals, but to model complete multi-channel EEG recordings. For this it will be important to further understand the impact of different design choices such as convolution size and up- and down-sampling techniques. In our experiments we noted a strong influence of convolutional size onto which frequency ranges are correctly expressed by the generator. Furthermore, we are currently applying our models to a large sample of EEG recordings from different subjects and will evaluate the quality of produced signals by an ensemble of medical experts.

In summary, EEG-GANs open up the possibility for new applications, not only limited to data augmentation, but e.g. also spatial or temporal super-sampling \cite{Corley2018} or restoration of corrupted signals. The possibility to generate signals of a certain class and/or with specific properties may also open a new avenue for research into the underlying structure of brain signals.



\end{document}